\documentclass[journal]{IEEEtran}

\usepackage{cite}
\usepackage[numbers,sort&compress]{natbib}
\usepackage{graphicx}
\usepackage{caption}

\ifCLASSINFOpdf
\else
\fi

\begin{document}

\title{Agent-Based Anti-Jamming Techniques for UAV Communications in Adversarial Environments: A Comprehensive Survey}

\author{Jingpu Yang$^{a}$,
         Mingxuan Cui$^{b}$
        Hang Zhang$^{a}$,
        Fengxian Ji$^{b}$,
        Zhengzhao Lai$^{b}$,
        and Yufeng Wang$^{a, \dagger}$%
\thanks{$^{a}$Jingpu Yang, Hang Zhang, and Yufeng Wang are with Beihang University, Beijing, 100191, China (e-mail: jingpuyang290@gmail.com; zhangh102@buaa.edu.cn; wyfeng@buaa.edu.cn).}%
\thanks{$^{b}$Fengxian Ji, Zhengzhao Lai and Mingxuan Cui with Northeastern University, Shenyang, Liaoning Province, 110004, China (e-mail: jifengxian123@outlook.com; pipixia688@gmail.com; mingxuancui3@gmail.com).}%
\thanks{$^{*}$Equal Contribution}%
\thanks{$^{\dagger}$Corresponding Author: Yufeng Wang.}
}

\maketitle
\begin{abstract}
Unmanned Aerial Vehicle communications are encountering increasingly severe multi-source interference challenges in dynamic adversarial environments, which impose higher demands on their reliability and resilience. To address these challenges, agent-based autonomous anti-jamming techniques have emerged as a crucial research direction. This paper presents a comprehensive survey that first formalizes the concept of intelligent anti-jamming agents for UAV communications and establishes a closed-loop decision-making framework centered on the "Perception-Decision-Action" (P-D-A) paradigm. Within this framework, we systematically review key technologies at each stage, with particular emphasis on employing game theory to model UAV-jammer interactions and integrating reinforcement learning-based intelligent algorithms to derive adaptive anti-jamming strategies. Furthermore, we discuss potential limitations of current approaches, identify critical engineering challenges, and outline promising future research directions, aiming to provide valuable references for developing more intelligent and robust anti-jamming communication systems for UAVs.

\end{abstract}


\IEEEpeerreviewmaketitle

\section{Introduction}
\IEEEPARstart{T}{he} rapid advancement of Unmanned Aerial Vehicle (UAV) technology leads to its increasingly widespread and in-depth applications in critical domains such as military reconnaissance \cite{military,mypaper}, disaster response \cite{disaster}, communication relay \cite{communication_relay}, intelligent transportation \cite{Transportation}, and future integrated space-air-ground networks \cite{Modlink,Integrated_Networks}. This results in an unprecedented reliance on stable, reliable, and efficient wireless communication links \cite{Multichannel_Aided}. However, the open and shared nature of wireless channels, coupled with UAVs' inherent high mobility, platform resource constraints, and complex deployment scenarios, makes them highly susceptible to various forms of interference in increasingly severe and dynamic adversarial electromagnetic environments. Effectively enhancing the anti-interference resilience of UAV communication systems in these environments is therefore a critical challenge. In particular, developing defensive mechanisms with autonomous decision-making capabilities is a challenging scientific and technological problem that requires urgent attention.

Due to the open and broadcast nature of wireless links, coupled with the high mobility of UAVs and increasingly complex electromagnetic environments, UAV communications are highly vulnerable to various forms of interference threats\cite{jamming_cancellation,UAV_jamming}. These threats encompass not only intentional malicious jamming initiated by adversaries to suppress or deceive but also mutual interference arising from spectrum sharing within UAV swarms, as well as environmental interference caused by complex terrain and atmospheric effects\cite{Electromagnetic,T-PAMI}. Although countermeasures such as frequency agility and beamforming \cite{MIMO_1,MIMO} exist, effectively applying these techniques in highly dynamic and information-incomplete real-world adversarial scenarios remains challenging. Consequently, enhancing the autonomous anti-jamming decision-making capability of UAV has emerged as a critical imperative to ensure their reliable operation.

Confronted with the aforementioned challenges, traditional anti-jamming methods, which rely on predefined rules or simplistic parameter adjustments, are proving increasingly inadequate. Consequently, the development of intelligent anti-jamming decision-making mechanisms, endowed with capabilities for autonomous learning, environmental adaptation, and intelligent game-theoretic interactions, has emerged as a pressing research imperative. To this end, the concept of the "Communication Anti-Jamming Intelligent Agent" has been introduced, which conceptualizes an UAV as an intelligent entity capable of proactively perceiving the external electromagnetic environment, making intelligent decisions aligned with its objectives and strategies, and autonomously executing corresponding anti-jamming actions. Key to empowering such agents are advanced artificial intelligence techniques; notably, Reinforcement Learning (RL) provides the core framework for policy optimization, Deep Learning (DL) offers powerful tools for complex environmental representation and function approximation, and Game Theory supplies the theoretical foundation for modeling adversarial and cooperative behaviors among multiple agents. The synergistic integration of these technologies is propelling UAV anti-jamming decision-making towards advanced levels of intelligence. To systematically understand and articulate the decision-making process of UAV anti-jamming intelligent agents, this review will focus on its core "Perception (P) - Decision-making (D) - Action Execution (A)" (P-D-A) closed-loop framework. Within this P-D-A framework, the perception phase is responsible for the accurate and efficient acquisition of diverse jamming-related information from complex environments. The decision-making phase, serving as the agent's cognitive core, leverages intelligent algorithms to generate optimal or near-optimal anti-jamming strategies based on perceptual inputs and predefined objectives. Finally, the action execution phase translates these strategic decisions into tangible and effective anti-jamming measures executable by the UAV platform.

In contrast to existing survey papers, this article presents and discusses the anti-jamming schemes from the perspective of agents. Compared with the existing survey papers, The main contributions of this article are as follows:
\begin{itemize}
  \item We explicitly define the concept of the UAV communication anti-jamming intelligent agent and systematically summarize its decision-making model based on the "Perception-Decision-Action" (P-D-A) closed-loop framework.
\item Centered on this P-D-A model, we outline the key technological challenges and representative intelligent solutions for each stage of UAV swarm communications. Particular emphasis is placed on elucidating how game theory can be employed to model the interactions between UAVs and jammers. Furthermore, we systematically review how other advanced strategies, such as reinforcement learning and deep learning, can be integrated to solve these game models.
\item We critically analyze the practical advantages and potential limitations of anti-jamming strategies that are based on game-theoretic modeling and solved using intelligent algorithms. Additionally, we explore key open challenges and promising future research directions in the domain of agent-based UAV anti-jamming communications.
\end{itemize}

\section{Background of Anti-Jamming Communications}
\subsection{Interference Types and Characteristics}
\begin{figure*}[!htbp]
    \centering
    \includegraphics[trim=5cm 6cm 4cm 6cm, clip, width=1\textwidth]{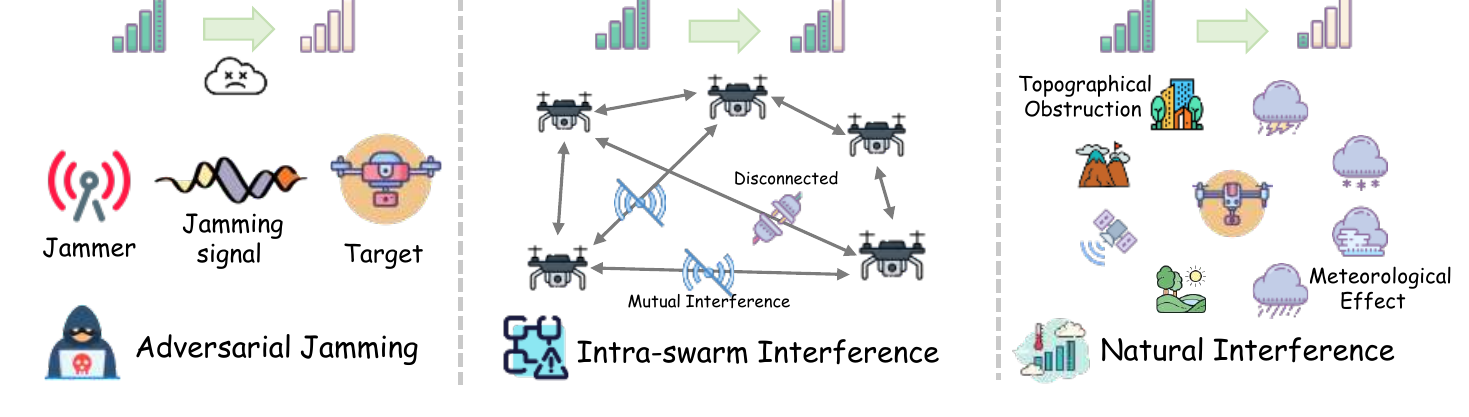}
    \caption{\textbf{The three primary categories of interference encountered in UAV communication systems.} }
    \label{fig:jamming}
\end{figure*}

Wireless communication systems, owing to their inherently open and shared nature, are highly susceptible to multiple forms of interference. This vulnerability is particularly pronounced in complex and dynamic UAV communication scenarios. Interference sources encompass not only malicious jamming initiated by adversarial actors but also mutual interference arising among legitimate users due to the scarcity of spectrum resources\cite{xu2018interference}. Furthermore, interference can originate from natural environmental factors\cite{he2019anti}(see Fig. \ref{fig:jamming}). Understanding the types and characteristics of interference is fundamental to designing effective anti-jamming strategies \cite{xie2021research}. This section, therefore, details different interference types and their primary characteristics.

\subsubsection{Adversarial Jamming}
Malicious interference specifically refers to wireless signals intentionally transmitted by adversarial entities or malicious users with the objective of disrupting, degrading, or denying targeted wireless communications\cite{he2016cyber}. Examples pertinent to UAVs include their control links, data transmission links, or navigation signal reception\cite{basan2022analysis}. UAV systems operating within complex electromagnetic environments are particularly vulnerable to such interference\cite{peng2019anti}. This vulnerability stems from the inherent openness of their communication links coupled with their critical dependence on stable control and precise navigation. Based on the interferer's mode of operation and intended effect, malicious interference is primarily classified into two fundamental categories: Suppressive Jamming and Deceptive Jamming\cite{kang2018reinforcement}.

\noindent\textbf{Suppressive Jamming}
The primary objective of suppressive jamming is to severely degrade the signal-to-interference-plus-noise ratio (SINR) at the UAV receiver by transmitting high-power radio frequency (RF) energy, thereby submerging the data signal under intense interference and rendering the UAV incapable of correct demodulation, ultimately disrupting its communication capabilities\cite{Suppressive}. Among diverse suppression techniques, broadband noise jamming achieves large-scale communication blockage by emitting high-power noise across a wide frequency spectrum\cite{broadband,broadband_1}, whereas narrowband jamming operates with higher precision by concentrating high-power interference on specific critical frequencies of UAV communications, creating an extreme interference field within the targeted narrowband to selectively paralyze particular links\cite{narrowband}. Sweep jamming employs frequency-agile interference signals that periodically scan potential UAV operational bands\cite{Sweep}, while pseudo-random sequence jamming transmits low-power signals with specific spreading code characteristics to disrupt despreading and signal processing at the receiver\cite{sequence}. All these suppression methods directly target the physical-layer reception of UAVs through distinct energy projection strategies, with the ultimate goal of severing their information exchange channels with external systems.

\noindent\textbf{Deceptive Jamming}
Deceptive jamming strategies fundamentally differ from direct energy suppression, as they employ sophisticated emulation of legitimate communication behaviors or signal characteristics to deceive unmanned aerial systems (UAS) into executing unintended and potentially harmful decisions \cite{Deceptive, Deceptive_1}. The primary threat of such interference lies in its high concealment and exploitation of logical vulnerabilities within communication protocol stacks. A prevalent form is spoofing attacks, wherein jammers simulate authentic signal sources by broadcasting structurally correct but navigationally erroneous signals, thereby compelling UAS navigation systems to compute false positioning, navigation, and timing (PNT) data. This can result in severe deviations from intended flight paths, unauthorized entry into restricted zones, or even complete loss of control \cite{li2018design, mendes2018effects, sathaye2022experimental, ma2024detection, majidi2018new, horton2018development}. Another archetypal deceptive technique involves replay attacks, where adversaries intercept and retransmit previously valid communication packets to exploit verification logic flaws, gaining unauthorized access or forcing systems to accept obsolete information \cite{xia2022identity, jan2022mutual, sen2025securing, alkatheiri2022lightweight}. Furthermore, redirection jamming—though not generating false content directly—manipulates signal propagation or introduces specific interference to induce erroneous source attribution by UAS receivers, potentially causing them to trust and execute commands from unauthorized entities \cite{sayeed2020safeguarding, guo2019position}. These "information deception" methodologies directly compromise the decision-making logic and operational integrity of UAS, presenting more complex and severe security challenges for autonomous or semi-autonomous platforms reliant on precise commands and trustworthy data.

\subsubsection{Interference in Inter-UAV Communications}
In UAV networks, particularly within dense swarm deployments, a critical challenge is mutual interference from internal communication activities among legitimate agents, distinct from external malicious attacks \cite{lim2024uav}. This arises from spectrum resource scarcity and escalating UAV communication demands, compelling multiple UAVs to use overlapping frequency bands for data sharing, collaborative control, or ground station communication \cite{wei2025integrated,javed2023communication,wu2022joint}. Spatial proximity in dense operations \cite{xu2021communication} and the swarms' high mobility and dynamically changing topology \cite{song2025slot} exacerbate this issue, leading to random, time-varying internal interference where the cumulative effect of signals from multiple neighboring UAVs can severely degrade performance at a receiving node.

This internal interference significantly degrades the SINR at the receiver, leading to diminished data rates, elevated Bit Error Rates (BER), and potential communication interruptions, thus directly jeopardizing UAV swarm mission reliability and efficiency\cite{lim2024uav}. Accurately distinguishing this internal interference from external malicious jamming in complex electromagnetic environments is also challenging. Addressing this requires advanced interference management techniques, such as optimizing resource allocation, frequency hopping, employing spatial domain methods like multi-antenna systems and beamforming, or managing UAV trajectories\cite{bui2023joint}. More sophisticated strategies involve using game-theoretic tools like graph or hypergraph games for modeling interference relationships, or applying collaborative Multi-Agent Reinforcement Learning (MARL) for the swarm to autonomously learn and cooperatively execute anti-interference policies \cite{hu2024evolutionary}.

\subsubsection{Environmental Interference}
In addition to anthropogenic interference, UAV communication systems inevitably face challenges imposed by a multitude of natural environmental factors. These sources of natural interference, originating from the properties and variations within the physical environment, primarily encompass atmospheric and meteorological effects, geographical constraints, and space environment dynamics. Specifically, ionospheric disturbances can adversely impact high-frequency (HF) communications and Global Navigation Satellite System (GNSS) signals, causing signal attenuation, phase fluctuations, and multipath propagation, thereby directly affecting UAV positioning accuracy and long-range communication efficacy\cite{ZHOU20244608}. Concurrently, meteorological phenomena such as rainfall, heavy snow, dense fog, dust storms, and lightning strikes significantly degrade signal quality through mechanisms like scattering, absorption, or the generation of potent electromagnetic pulses, posing a considerable threat particularly to UAV data links and satellite relay links operating at microwave and higher frequency bands\cite{ev}. Furthermore, UAV operations in complex terrains—including low-altitude, urban, or mountainous environments—are subject to signal degradation caused by physical obstructions like buildings, mountains, and vegetation, which induce signal blockage, reflection, and diffraction, leading to coverage dead zones and pronounced multipath effects; ground reflections can also contribute additional interference\cite{drones7070423}. For UAVs reliant on satellite links for Beyond-Line-of-Sight (BLOS) control or data backhaul, space weather events such as solar activity-induced geomagnetic disturbances and ionospheric scintillation present another layer of challenge by potentially disrupting these critical satellite communication links, especially in high-latitude regions\cite{ZHAO20242492}. 

Comprehensively considered, the natural interference environment inhabited by UAVs is characterized by significant multidimensional spatio-temporal dynamics. Meteorological conditions, ionospheric states, and even the UAV's own position and attitude are in constant flux, imparting stochastic and time-variant properties to the experienced interference. Furthermore, the accurate prediction and modeling of these natural environmental factors often present considerable challenges, confronting communication system design with the inherent difficulty of incomplete information\cite{ISHII2024}. Compounding this complexity, natural interference frequently coexists and interacts with mutual interference generated internally within UAV swarms, as well as potentially present external malicious interference. This superposition of diverse interference sources culminates in an exceptionally complex electromagnetic environment for UAV operations \cite{fan2024performance}. Consequently, the design of robust anti-interference communication systems capable of stable and reliable operation within such dynamic environments—marked by informational uncertainty and the confluence of multiple interference origins—represents a pivotal research direction within the field of unmanned aerial systems.

\subsection{Classification of Anti-jamming Strategies}
\begin{figure*}[!htbp]
    \centering
    \includegraphics[trim=5cm 6cm 4cm 3.5cm, clip, width=1\textwidth]{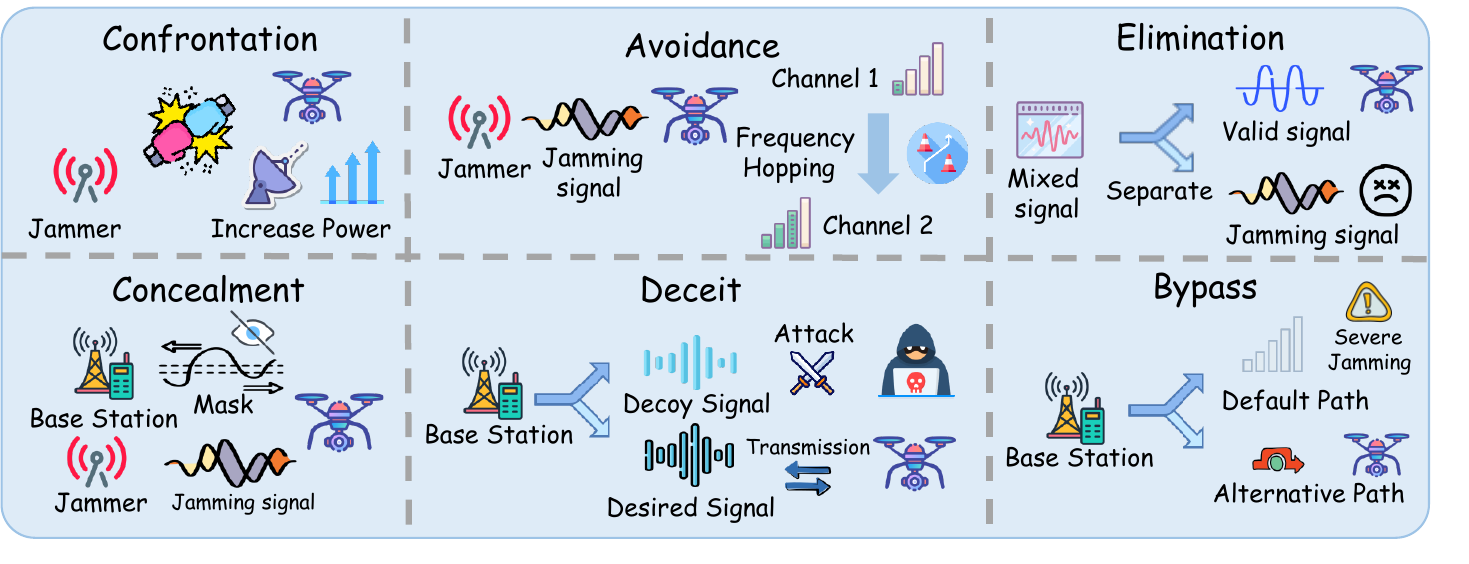}
    \caption{\textbf{The six Anti-Jamming strategies for UAVs.} }
    \label{fig:six}
\end{figure*}

In the field of UAV wireless communications, researchers have proposes various anti-jamming strategies to counter increasingly complex and intelligent interference threats. These strategies can be broadly categorized into six types based on their core concepts and implementation methods: adversarial, avoidance, cancellation, covertness, deception, and circumvention(see Fig. \ref{fig:six}). Each of these strategies has a distinct focus and is suitable for different interference environments and communication requirements. The selection and combination of these strategies are particularly critical for UAV platforms, which are often resource-constrained and highly mobile.

\textbf{Confrontation:} The core idea of the confrontation strategy is to suppress interfering signals by enhancing the strength of the desired signal. The most direct method is to increase the transmission power, thereby improving the Signal-to-Interference-plus-Noise Ratio (SINR) at the receiver \cite{ma2024intelligent,shi2019power}. \cite{distributed,distributed_1} models a Stackelberg game to depict the adversarial relationship between the UAV and the jammer.

\textbf{Avoidance:} The core idea of the avoidance strategy is to evade the impact range of interference. Common avoidance techniques include frequency hopping\cite{FHTH,Link-SAC}, which involves switching to other unjammed channels when interference occurs, and spatial isolation, such as utilizing multi-antenna or beamforming technologies on UAVs to adjust signal direction and steer clear of the incoming direction of interference signals. \cite{CFHA} proposes a lightweight method combining frequency hopping and chaotic mapping, effectively mitigating interference attacks by leveraging the unpredictability and randomness of chaotic mapping to create secure frequency hopping sequences. LRAJ \cite{LRAJ} actively dodges jammed frequencies through frequency replacement and adaptive frequency hopping (AFH).

\textbf{Elimination:} This strategy focuses on mitigating the impact of interfering signals at the receiver end through signal processing techniques. Commonly employed methods include adaptive filtering and blind source separation \cite{PID}. \cite{jamming_cancellation} developed a practical multi-channel auxiliary interference cancellation method to enable secure UAV communication, capable of simultaneously achieving timing/frequency synchronization and interference suppression. \cite{MIMO} leverages Multiple-Input Multiple-Output (MIMO) technology to alleviate anti-jamming challenges in UAVs by capitalizing on its advantages in directionality, diversity, multiplexing, and interference cancellation (BIJC) schemes.

\textbf{Concealment:} The objective of this strategy is to reduce the probability of the communication signal being detected by a jammer, thereby making it difficult for the jammer to ascertain the existence of communication activity. Direct-Sequence Spread Spectrum (DSSS) techniques can enable the transmission of signals below the noise floor. \cite{spread_spectrum} achieved a three-orders-of-magnitude reduction in interception probability by establishing a covert channel for transmitting dynamic spreading codes.

\textbf{Deceit:} This strategy constitutes an active defense mechanism where false information is transmitted or misleading actions are undertaken to induce erroneous decision-making by a jammer or to redirect its attacks towards invalid targets, thereby safeguarding the transmission of legitimate information \cite{Deceit}. Examples include deploying an auxiliary transceiver to emit decoy signals specifically to attract jamming resources, or having the transmitter intentionally adopt suboptimal operational behaviors to mislead the predictive models employed by an intelligent jammer. Deceit strategies are considered particularly potent against intelligent or reactive jammers.


\textbf{Bypass:} This strategy represents a network-layer anti-jamming mechanism, distinct from traditional physical layer or MAC layer approaches. It leverages routing diversity within the network. When a specific area experiences severe interference, alternative paths are selected to bypass the affected region, thereby re-establishing the communication link. \cite{homogeneous_sensors} utilizes frequency conversion to circumvent the GNSS navigation band, achieving physical layer aggregation of signals from multiple UAVs.

\subsection{Analysis of Challenges in Anti-Jamming Decision Making}
The formulation of anti-jamming decision-making for UAVs executing communication missions confronts severe and multifaceted challenges. These challenges are deeply rooted in the unique application scenarios, platform constraints, and the increasingly complex electromagnetic countermeasure environment inherent to UAV operations. Consequently, the design of genuinely effective and reliable anti-jamming strategies necessitates a profound understanding and adept management of these difficulties. Specifically, the principal challenges manifest in the following aspects:


\textbf{Highly Dynamic and Complex Interference Environment Challenge:} This represents one of the most significant challenges in UAV communication. The inherent high-speed, three-dimensional mobility of UAVs—characterized by rapid changes in position, attitude, and swarm formations—induces severe and fast variations in wireless channel characteristics. Consequently, the acquisition and prediction of Channel State Information (CSI) become exceptionally challenging. Concurrently, interference sources, potentially including highly mobile entities such as adversarial UAVs, can exhibit rapid changes in location, power, directionality, and strategy, further exacerbating the environmental dynamics. The complexity is compounded by the dynamic nature of network topologies, particularly in ad-hoc networking (MANETs) and swarm scenarios, as well as evolving mission requirements\cite{Topology}. Especially in dense deployment scenarios like UAV swarms, individual UAVs must contend not only with external malicious interference but also with significant mutual interference arising from spectrum sharing within the swarm itself. The superposition and coupling of these internal and external interference types render the identification of interference sources, estimation of interference channels, and implementation of effective interference suppression techniques extraordinarily difficult. This places stringent demands on cooperative communication protocols, resource management algorithms, and distributed decision-making processes.

\textbf{Information Uncertainty and Adversarial Intelligence Challenge:} UAVs face significant information asymmetry and escalating threats when countering adversarial interference. Due to the non-cooperative nature of adversaries, limitations in detection capabilities, and strict constraints on UAV platforms regarding sensors, payload, computation, and power consumption, it is challenging for UAVs to acquire complete and accurate real-time information about the interference (e.g., type, parameters, intent, location). They may even encounter entirely novel interference patterns. Decision-making based on incomplete, imprecise, or even deceptive information is highly prone to leading to strategy failure. Concurrently, adversaries employing interference are transitioning from simple power suppression tactics towards more intelligent approaches\cite{challenge}. Intelligent jammers can learn UAV communication behaviors and flight patterns, dynamically adjust their jamming strategies for precise targeting or deceptive interference, and even launch attacks against specific protocols. This emerging "intelligence versus intelligent interference" dynamic necessitates that UAVs possess enhanced situational awareness and sophisticated intelligent game-theoretic capabilities.

\textbf{Robust and Efficient Decision Algorithm Design Challenge:} Designing effective anti-jamming decision-making algorithms under dynamic, uncertain, and extreme conditions presents significant challenges. The algorithm must not only rapidly generate valid judgments with strong adaptability and generalization capabilities against unknown interference but also achieve a delicate balance among multiple competing performance metrics, including real-time responsiveness, stability, and convergence properties \cite{Topology}. Furthermore, it must simultaneously address complex coexisting internal and external disturbances while coordinating potentially conflicting objectives such as communication performance, energy efficiency, and covertness under stringent resource constraints of UAV platforms. These stringent requirements render conventional optimization methods limited due to their reliance on precise modeling, whereas learning-based approaches like reinforcement learning still face fundamental challenges in sample efficiency, safety guarantees, and theoretical robustness.

\section{Definition of UAV Anti-Jamming Agents}

\begin{figure*}[!htbp]
    \centering
    \includegraphics[trim=5cm 5cm 6cm 3cm, clip, width=1\textwidth]{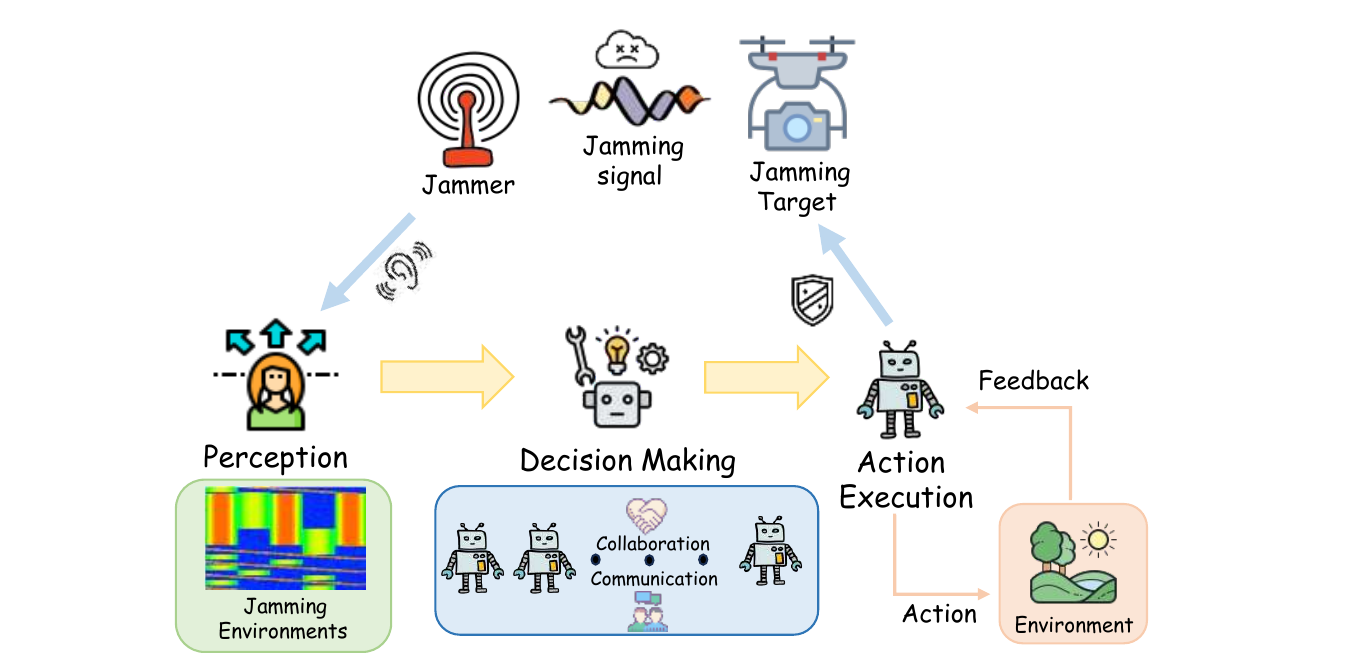}
    \captionsetup{justification=centering}
    \caption{\textbf{The decision-making process for anti-jamming in UAV communications} }
    \label{fig:flowchart}
\end{figure*}

An \emph{agent} is an entity or system capable of perceiving its environment, making autonomous decisions based on such perceptions, and executing actions to achieve predefined objectives. In general, an agent possesses three core capabilities:

\begin{itemize}
    \item \textbf{Perception:} The ability to gather and interpret real-time information from the environment using sensors or other sensing devices.
    \item \textbf{Decision-making:} The capacity to analyze environmental states based on perceived information and predetermined goals, employing reasoning or learning algorithms to determine optimal or near-optimal actions.
    \item \textbf{Action:} The capability to execute decisions through actuators or other means, actively influencing or altering the environment to accomplish specific goals.
\end{itemize}
Extending the general concept of intelligent agents, this study concentrates on a specialized application: communication anti-jamming agents integrated into UAV systems. These UAV-deployed agents function as autonomous decision-making entities characterized by their capacity for real-time perception of the electromagnetic environment, the identification and analysis of interference patterns, and the execution of dynamic anti-jamming decisions and actions. By leveraging onboard sensors and processing units within a closed-loop "Perception-Decision-Action" framework(see Fig. \ref{fig:flowchart}), these agents can adaptively modify communication protocols or even flight parameters. This dynamic adaptation occurs in response to complex, potentially adversarial electromagnetic conditions, with the objective of preserving the stability, reliability, and overall effectiveness of essential communication links.

A UAV anti-jamming intelligent agent must first exhibit real-time perceptual capabilities within complex electromagnetic environments. This involves utilizing onboard sensors to continuously detect and monitor the electromagnetic spectrum proximate to the flight path. Concurrently, the agent requires the capacity for rapid identification and analysis of non-stationary interference patterns that affect UAV communications. This capability enables the swift recognition, classification, and assessment of detected interference signals to ascertain their specific impact on the integrity of the communication link. Ultimately, leveraging the perceived signals and analytical outcomes, the intelligent agent optimizes its operational parameters and learns adaptive anti-jamming strategies. This facilitates the execution of comprehensive countermeasures, including dynamic switching of communication frequency bands, intelligent control of transmission power, beamforming, and adjustments to the flight trajectory.

Conclusively, intelligent agents designed for anti-jamming communication in UAVs, operating on the "Perception-Decision-Action" cycle and enabled by autonomous intelligence, possess the capability to efficiently perceive and adapt to complex, dynamic electromagnetic environments. Through the dynamic adjustment of communication strategies and possible physical avoidance maneuvers, these agents, when deployed on UAVs, markedly improve the robustness and transmission efficiency of the communication systems, particularly when facing severe interference or non-cooperative scenarios. This furnishes a solid foundation for advancing more intelligent and dependable UAV communication networks and application frameworks.

\section{Theoretical Foundations and Core Mechanisms of UAVs Anti-Jamming Intelligent Agents}

The decision-making process of intelligent agents for anti-jamming in UAV swarm communications is an autonomous collaborative system centered around the "Perception-Decision-Action" closed-loop. UAVs must contend with dynamic and complex electromagnetic environments, as well as constraints of real-time performance and security. Leveraging their distributed collaborative characteristics, the swarm achieves rapid and precise interference identification, localization, and threat analysis through multi-node shared perception. Subsequently, based on shared situational awareness and mission requirements, optimal or suboptimal anti-jamming strategies are formulated in parallel and executed cooperatively.

\subsection{Perception Layer}
The perception layer serves as the interactive front-end between the UAV's anti-jamming intelligence and the complex electromagnetic environment. Its core function is to achieve real-time and accurate situational awareness of the battlefield to support intelligent decision-making. Perception in UAV swarms is confronted with challenges such as time-varying channels induced by high-speed maneuvers, limited platform resources, and communication links susceptible to multiple types of interference. Furthermore, the swarm must address issues related to collaborative perception and intra-swarm interference. Consequently, the perception layer is required to be efficient, robust, and possess low overhead. Its primary tasks encompass spectrum sensing, interference identification and feature extraction, and jammer localization.
Regarding spectrum sensing, traditional methods such as energy detection and matched filtering struggle to cope with complex environments due to their sensitivity to low signal-to-noise ratios (SNR) and unknown signal types. Time-Frequency Analysis (TFA) can effectively process non-stationary signals to identify sweeping or frequency-hopping patterns, while Hidden Markov Models (HMMs) are suitable for predicting dynamic spectrum occupancy by modeling state transitions. As a specific implementation of active perception to support anti-interference decision-making, \cite{RVR} proposes utilizing an airborne monitoring platform to perceive electromagnetic interference (EMI) characteristics  and employing Relevance Vector Machine Regression (RVR)  to predict interference effect thresholds, thereby guiding adaptive anti-interference measures. To address the high mobility of UAVs and rapid spectral changes, methods combining HMM, TFA, and historical sequence analysis can help track channel states and predict interference frequency bands for active avoidance.

Regarding jammer localization and collaborative perception, an individual UAV possesses a limited perception range, rendering it challenging to accurately assess dynamically varying interference signals from diverse directions. This poses inherent limitations, particularly in localizing interference sources. For instance, a single node might only perform coarse estimations based on Angle of Arrival (AoA) data or utilize its own mobility for "motion-based perceptive localization," inferring the interference source's direction from measurements acquired at different positions.

In contrast, a UAV swarm can leverage the advantages of its multi-node architecture and wide spatial distribution to achieve perception capabilities far exceeding those of a single UAV. By enabling individual UAV nodes to share their detected environmental information, the swarm can synthesize the information gain afforded by spatial diversity and perform data fusion. This facilitates more rapid and precise localization and identification of interference sources, thereby overcoming the blind spots and errors associated with single-point sensing. Furthermore, advanced techniques, such as the utilization of Graph Convolutional Networks (GCNs) to predict the location and intensity of interference regions \cite{GCN}, offer new avenues for swarm collaborative perception and subsequent avoidance decision-making.

Concurrently, accurate situational awareness is a fundamental prerequisite for effective network resource optimization. For instance, \cite{Topology} proposes a topology optimization method to maximize the minimum network capacity by jointly designing the UAVs' transmission paths and power. This type of optimization is highly dependent on the accurate perception of current channel conditions, network topology, and potential interference distribution.

\subsection{Decision-making Layer}
The decision-making layer is central to anti-jamming in UAV swarms; it leverages perceived information, self-status, and mission objectives to generate efficient and robust collaborative anti-jamming strategies. Swarm decision-making must address challenges such as dynamic environments, limited resources, multi-dimensional coordination, and incomplete information. Consequently, decision-making must balance multiple objectives, including communication performance, energy efficiency, and security, and output multi-dimensional cooperative anti-jamming actions encompassing aspects such as spectrum, power, beamforming, and flight trajectory.

Collaborative decision-making in UAV swarms is based on fused perceptual information, achieving superior global or local anti-jamming effectiveness through multi-agent collaboration. To this end, Multi-Agent Reinforcement Learning has emerged as a core technology. The key to MARL lies in achieving effective collaboration. For instance, \cite{MATD3,effTD3} promotes cooperation and addresses uncertainty by designing team rewards for each UAV agent. In terms of training paradigms, Centralized Training with Decentralized Execution (CTDE) is a common strategy. As an example, \cite{CMJD} employs the CTDE framework based on the classic reinforcement learning algorithm Actor-Critic (A2C) to collaboratively counter reactive jamming.

For different anti-jamming decision-making problems, MARL exhibits a diverse range of solutions. When addressing discrete collaborative actions, such as joint channel selection, value-function-based methods are applied. For example, in \cite{CMDA}, UAVs within a swarm independently update their local Q-networks based on local observations of the environment and a common reward to learn collaborative policies. For collaborative tasks requiring continuous control, such as swarm trajectory planning and joint resource regulation, policy-optimization-based MARL algorithms are more effective. For instance, \cite{MADDPG_path_planning} utilizes the Multi-Agent Deep Deterministic Policy Gradient (MADDPG) algorithm to plan robust anti-jamming flight paths for swarms under complex constraints.

More sophisticated MARL strategies further enhance the intelligent decision-making capabilities of swarms to cope with dynamic adversarial situations. For example, \cite{HMAPPO} decouples hierarchical mixed game problems into several sub-Markov games and updates actor-critic networks at different time scales. Beyond direct anti-jamming confrontations, MARL is also employed to optimize specific system objectives. For instance, \cite{energy_efficient_framework} designs a distributed reinforcement learning algorithm that not only strives to improve communication quality when under jamming attacks but also endeavors to reduce the total energy consumption of the entire network. CMAA \cite{CMAA} enables users to collaboratively select channels by sharing their respective learning experiences and jointly evaluating the merits of different joint channel selection strategies.

Facing the challenge of decision-making complexity brought by large-scale swarms, Mean Field Game (MFG) theory \cite{MFG} offers an effective simplification pathway by approximating complex inter-agent interactions through an average effect. Furthermore, the realization of efficient collaborative decision-making also depends on the support of the underlying system architecture. For instance, \cite{Dyna-Q} proposes a network architecture that integrates UAV flight controllers with Software-Defined Network (SDN) controllers, providing system-level support for the implementation and deployment of swarm collaborative decision-making.

\subsection{Action Execution Layer:} 
The action execution layer serves as the ultimate phase for a UAV (Unmanned Aerial Vehicle) swarm to implement its anti-jamming strategy. It is responsible for precisely translating collaborative decisions into both the joint adjustment of physical layer communication parameters and coordinated maneuvering commands for the swarm. This process necessitates the effective management of various constraints for each UAV—including its dynamic characteristics, energy limitations, and flight safety parameters—and employs efficient, precise, and safe coordinated control to ensure the attainment of the swarm's overall anti-jamming objective.

The execution of anti-interference actions for UAV clusters critically relies on translating cooperative decisions into precise physical maneuvers. In terms of cooperative physical maneuvering and spatial planning, the cluster mitigates interference through collective path planning and formation control. For instance, \cite{APF} utilizes an artificial potential field model to guide multi-UAV collaborative path planning to evade interference sources. Concurrently, \cite{WD5PM,NT5D} employed deep reinforcement learning to jointly optimize UAV trajectories and user associations, aiming to alleviate the impact of interference on tasks. Diverging from DRL approaches, \cite{Joint_optimization} proposes an iterative algorithm based on Block Coordinate Descent (BCD) and Successive Convex Approximation (SCA) to jointly optimize UAV trajectories and cluster formation for interference mitigation. Furthermore, \cite{game_MADRL} applied game theory to optimize the spatial strategies of clusters in interference environments.

In terms of cooperative communication and resource regulation, the cluster can jointly adjust communication parameters to maintain link stability. For example, \cite{SLA} proposes a distributed cooperative anti-interference channel access method to improve decision convergence and ensure channel availability under interference. Similarly, \cite{GAED} utilizes a specific genetic algorithm (GAED) to determine the optimal frequency hopping speed and transmission power for transmitters; the precise execution of these parameters is crucial for enhancing communication covertness and anti-interference capabilities.

To achieve higher-level cluster-wide effectiveness and resilience, complex cooperative anti-interference actions are executed. For example, \cite{CETA} investigated the target association problem in cluster evolution, enhancing task flexibility and collaborative efficiency by dynamically partitioning and matching sub-clusters with anti-interference tasks. Meanwhile, \cite{homogeneous_sensors} utilizes multiple UAVs (with homogeneous sensors) to improve system elasticity and spatial degrees of freedom, thereby simultaneously achieving objectives such as anti-interference and cooperative navigation.

Ensuring the safe and efficient execution of cooperative actions is also crucial. The action execution layer must integrate security assurance mechanisms, such as Control Barrier Functions (CBF), to monitor in real-time and rectify actions that might violate safety constraints. To enhance response and optimization efficiency in complex environments, hierarchical management mechanisms are often adopted, decomposing tasks by time scales. For instance, in resource block-level optimization, techniques such as convex optimization, Successive Convex Approximation (SCA) \cite{SCA,SCA_1}, and Block Coordinate Descent (BCD) \cite{BCD} can be employed to cooperatively optimize internal cluster resources like power allocation. Model Predictive Control (MPC) is also introduced, utilizing proactive optimization to improve control precision and ensure link stability.


\section{Technical Aspects of Agent-Based UAV Anti-Jamming Strategies}

The core of agent-based UAV anti-jamming strategies lies in advanced modeling and learning algorithms for autonomous decision-making in complex electromagnetic environments. Game theory effectively analyzes adversarial interactions between UAVs, jammers, and intra-swarm competition, offering a theoretical framework. However, real-world scenarios are dynamic with incomplete information, where reinforcement learning excels by enabling strategy learning through trial-and-error without full prior jammer knowledge. Consequently, integrating game theory's interaction modeling with reinforcement learning's environmental learning and decision optimization—forming a game-theoretic learning framework—is a key direction for tackling intelligent UAV anti-jamming challenges.


\subsection{Game Formulation} 
In this framework, the primary step is precise game formulation. This requires an in-depth analysis of the specific characteristics of the UAV anti-jamming scenario to select the game model that best reflects its inherent interactive logic. Different game models are applied to characterize the complex interactions between UAVs and jammers, as well as within UAV swarms.

Stackelberg game models, due to their ability to capture leader-follower dynamics, are widely used in UAV anti-jamming. For instance, \cite{proposes_algorithm,Bayesian-Stackelberg} models the interaction between the UAV and the jammer as a Stackelberg game, converting it into a bi-matrix formulation to determine the UAV's optimal mixed strategy when facing interference. In this model, one party (the leader) acts first, and the other party (the follower) makes an optimal response after observing the leader's strategy. In \cite{annealing_algorithm}, an intelligent jammer acts as the leader, possessing the prerogative to act before the user, while the user, after learning the jammer's strategy, acts as a follower, adaptively adjusting transmission power to maximize their utility function. Similarly, \cite{HMAPPO} constructs a Stackelberg game between a UAV jammer (leader) and ground users (followers). \cite{beamforming} models satellites, UAVs, and Intelligent Reflecting Surfaces (IRS) as leaders, with the jammer as the follower, to describe their adversarial relationship. Furthermore, hierarchical game approaches also leverage Stackelberg models, such as the algorithm in \cite{propose}, which employs a Stackelberg game to address the UAV's position deployment problem, and the second layer utilizes a potential game for channel allocation.

Potential games and Markov games are employed to analyze convergence and equilibrium properties in multi-agent systems. \cite{MALQL} models the UAV anti-jamming problem as a local interaction Markov game and proved it to be an exact potential game with at least one Nash equilibrium. This type of formulation is useful for analyzing and designing distributed learning algorithms where UAVs can reach a system-level stable state through local information and interactions.

Differential games and pursuit-evasion games are suitable for describing dynamic adversarial processes where states evolve continuously over time. \cite{MFG} models an optimization problem as a differential game, where the dynamics of interference and energy consumption are models as the states of the participants, allowing for the analysis of optimal strategies in long-term confrontations. \cite{Isaacs} approximated a capacity optimization problem as a zero-sum pursuit-evasion game, utilizing Isaacs' equations for its solution, which is often used to analyze worst-case performance guarantees.

Other game models and cooperative approaches also provide solutions for specific scenarios. For example, to foster cooperation within a UAV swarm, each UAV in \cite{distributed,distributed_1} models a local altruistic game, considering its own utility and that of other UAVs, thereby proposing a distributed cooperative anti-jamming algorithm. CLASA \cite{CLASA}, on the other hand, constructed an "external adversarial - internal coordination" game model for anti-jamming spectrum access, distinguishing between the adversarial interaction with external jammers and the coordination of spectrum sharing within the UAV swarm.

\subsection{Intelligent Learning Algorithm} 
Following the game model's construction, the core task is designing intelligent learning algorithms to solve for the game equilibrium and obtain effective anti-jamming strategies that can be effectively executed in practical environments. Given that the environments encountered by UAVs are typically dynamic, with incomplete and unknown information, the algorithms must possess strong capabilities for environmental adaptation and autonomous decision-making. Based on the rewards obtained, the agent continuously adjusts its own strategy to maximize cumulative returns. Currently, various intelligent learning paradigms are applied in this domain; Reinforcement Learning is particularly prominent, while other techniques from Deep Learning also play significant roles.


\subsubsection{Q-learning and its Extensions}
Q-learning \cite{Q-learning} is a classical reinforcement learning method widely applied in anti-jamming communications. It enables agents to learn through interaction with dynamic environments, updating a state-action value function based on received rewards to optimize decision-making policies. Its primary advantage lies in the ability to learn directly from environmental interactions and adapt to dynamic jamming without requiring precise prior knowledge of interference patterns. ICADCSA \cite{ICADCSA} proposes a Q-learning-based distributed cooperative anti-jamming channel selection algorithm to address both malicious interference and co-channel interference problems.

To address more complex anti-jamming scenarios and decision-making problems, various advanced extensions of Q-learning have been developed. For instance, \cite{double_deepQ} proposes a dueling double deep Q-network (D3QN) multi-step learning algorithm aimed at solving complex coupled decision-making problems. Such advanced deep Q-network architectures, like D3QN, have been employed to determine optimal anti-jamming paths for UAVs\cite{D3QN, D3QN_1}. In the context of multi-agent and distributed systems, researchers have also made numerous explorations: MD-QL\cite{MD-QL} utilizes clustering methods, task allocation, and online learning techniques to enhance multi-user independent learning for avoiding external and internal interference while addressing the heterogeneous task requirements of UAV swarms; MALQL\cite{MALQL} implements anti-jamming communication based on cooperative multi-agent hierarchical Q-learning to reduce the dimensionality of the action space; PFSAC\cite{PFSAC} combines global and local models to customize personalized anti-jamming strategies for each UAV, significantly enhancing network performance in complex interference environments; and FDQN\cite{FDQN} integrates federated learning with Q-learning, enabling multiple UAVs to collaboratively optimize their channel selection and power allocation strategies in a distributed manner. Furthermore, other optimization methods have been proposes: DDQNR\cite{DDQNR} employs a D3QN-based relay method, where UAVs optimize their flight trajectory and transmission power during relay communication to dynamically adapt to and evade interference; Dyna-Q\cite{Dyna-Q} incorporates an environment model to cooperatively optimize UAV power allocation and trajectory planning to counter intelligent jamming attacks; and ETMAPPO\cite{ETMAPPO} utilizes beta operators for optimization search, aiming to achieve better global returns with fewer information dimensions, which is crucial for enhancing decision-making efficiency in complex environments.

\subsubsection{Deep Reinforcement Learning} While Q-learning is widely applied in the field of anti-jamming, it faces the "curse of dimensionality" when dealing with high-dimensional state spaces. This leads to diminished learning efficiency and difficulties in convergence. Deep Reinforcement Learning has emerged as a key technology to overcome this challenge. DRL effectively combines the powerful representation learning capabilities of Deep Neural Networks with the decision-making framework of reinforcement learning. By utilizing various DNN architectures to approximate or replace the Q-function or policy function inherent in traditional RL, DRL can directly process high-dimensional raw inputs. This enables the learning of complex anti-jamming strategies in dynamic and unknown interference environments.

The specific applications of DRL in the anti-jamming domain showcase its diverse solution methodologies and significant potential. Primarily, in the realm of interference perception and environmental representation, DRL can effectively extract critical information from complex inputs. For instance, CEN \cite{CEN} utilizes an interference feature map as the input to a Convolutional Neural Network (CNN) and employs a normalized multi-objective function value as the output to guide the training process. Similarly, the work in \cite{DNN} involved designing and training an online Signal-to-Noise Ratio (SNR) mapping deep neural network through supervised learning to encode the impact of jammers. Furthermore, \cite{GCN} leverages a Graph Convolutional Network (GCN) to predict the location and intensity of interference zones based on information collected from each UAV.

Secondly, based on environmental perception or by learning directly through interaction with the environment, DRL plays a pivotal role in dynamic resource optimization and intelligent decision-making. For instance, under conditions where the interference model and interference power are unknown, the approach in \cite{DDPG_CSI} utilizes DRL to dynamically adjust the transmission power of UAVs by estimating the effective jamming signal strength (EJSS) and using this estimation as an input. For more intricate joint optimization problems, \cite{HLASG-DDPG} employs a combination of Q-learning and stochastic learning automata for channel selection, further integrated with DDPG for power control. The DRUMC framework \cite{DRUMC} leverages DRL to adaptively select optimal power control strategies via independent Dueling Neural Networks, thereby effectively countering proactive interference and mitigating energy consumption without dependence on a specific interference model. Furthermore, Noisy-D3QN-PER \cite{Noisy-D3QN-PER} constructs robust transmission beams to combat jamming and eavesdropping through the joint optimization of base station power allocation and the discrete phase shifts of UAV-mounted Reconfigurable Intelligent Surfaces (RIS).

Moreover, researchers are actively exploring innovative DRL frameworks and learning paradigms to enhance anti-jamming performance. For example, the work in \cite{Decision_Transformers} employs an offline anti-jamming decision-making method based on Decision Transformers to develop more practical anti-jamming decision models. DMURP \cite{DMURP}, on the other hand, demonstrates the general capability of DRL to process and approximate objective functions using DNNs, thereby providing a flexible toolkit for addressing complex anti-jamming problems.

\subsubsection{Transfer Reinforcement Learning} 
Transfer reinforcement learning offers a promising technical approach to enhance the adaptability of UAVs in anti-jamming scenarios. The core objective is to effectively transfer and reuse knowledge acquired from source tasks or simulation environments to new or dynamically changing interference environments, thereby significantly improving learning efficiency and accelerating convergence to reliable anti-jamming strategies.

In the application of anti-jamming for UAVs, transfer reinforcement learning has demonstrated various effective practical approaches. For instance, MURP and DMURP \cite{MURP,DMURP} enhance sample efficiency and accelerate policy optimization by enabling UAVs to share learning experiences. Similarly, leveraging transfer learning to initialize the weights of Convolutional Neural Networks (CNNs) within DRL networks is a common acceleration technique. SHDRL \cite{SHDRL} adopts this method, achieving efficient initial learning through the transfer of CNN weights between agents. Addressing specific application needs, such as safeguarding the Quality of Experience (QoE) for UAV video transmission, SRL-AJ \cite{SRL-AJ} developed a secure transfer DRL anti-jamming scheme aimed at ensuring video QoE and reducing outage probability.

Furthermore, warm-start techniques, which involve transferring previously trained Q-values or experiences, are widely employed to expedite the learning of optimal policies \cite{REAR}. In scenarios where UAVs act as communication relays against intelligent jamming, transferring pre-trained Q-values can accelerate the formulation of the optimal relaying strategy \cite{PHC}. Fast DQN \cite{Fast_DQN} proposes a frequency-space two-dimensional anti-jamming scheme based on DQN and utilizes warm-start experience transfer to more effectively counter jamming attacks. More generalized knowledge transfer, such as migrating Q-values or action-state pair information across regions or tasks, is also crucial for UAVs requiring rapid adaptation to diverse environments. For example, AQLA \cite{AQLA} introduced a multi-region transfer RL scheme for channel selection via Q-value transfer, while TGACT \cite{TGACT} proposes a transfer Actor-Critic scheme in cognitive radio networks that migrates action and state information to optimize channel selection.

\subsubsection{Other Reinforcement Learning-Based Methods}
Beyond deep reinforcement learning (DRL) and transfer reinforcement learning (TRL), several other advanced reinforcement learning paradigms, such as meta-reinforcement learning (Meta-RL), federated reinforcement learning (FRL), knowledge-based reinforcement learning (KBRL), and multi-armed bandits (MAB), also demonstrate significant potential in addressing complex anti-jamming challenges in wireless communications within the UAV domain. These approaches offer new avenues for UAVs to achieve intelligent decision-making in dynamic and adversarial environments.

Meta-RL aims to enable agents to rapidly adapt to new tasks or dynamically changing environments based on limited experience. For instance, \cite{MMFQ} combined meta-learning with mean-field Q-learning to provide a solution for rapid adaptation problems under dynamic jamming. On the other hand, KBRL enhances learning efficiency and performance by integrating domain knowledge, such as UAV dynamics and communication principles, into the learning process. The work in \cite{KBRL} achieved joint optimization of flight control and power allocation to counteract smart jamming by constructing a virtual environment embedded with domain knowledge for pre-training.

FRL offers an effective distributed learning framework for collaborative anti-jamming in UAV swarms. It allows multiple UAVs to cooperatively train a shared anti-jamming model while preserving data privacy, making it particularly suitable for Flying Ad Hoc Networks (FANETs). For instance, UAVs can independently train models locally and then aggregate their parameters to collectively counter smart jamming \cite{PFSAC, FDQN}. Alternatively, they can collaboratively share jamming knowledge via federated learning to avoid jammed areas, thereby optimizing the interference avoidance performance of the entire network \cite{AFRL}.

Furthermore, MAB as tools for managing the exploration-exploitation trade-off, are frequently employed in scenarios requiring online learning of optimal actions, such as channel selection \cite{UAV-CAV} or power control. Algorithms like EXP3 \cite{EXP3} and UCB are used to learn optimal strategies under unknown interference. KL-UCB \cite{UCB,KL-UCB} addresses the MAB problem by selecting the arm with the highest index based on the Kullback-Leibler divergence when choosing actions.


\begin{figure*}[!htbp]
    \centering
    \includegraphics[trim=4cm 4.5cm 7.5cm 3cm, clip, width=1\textwidth]{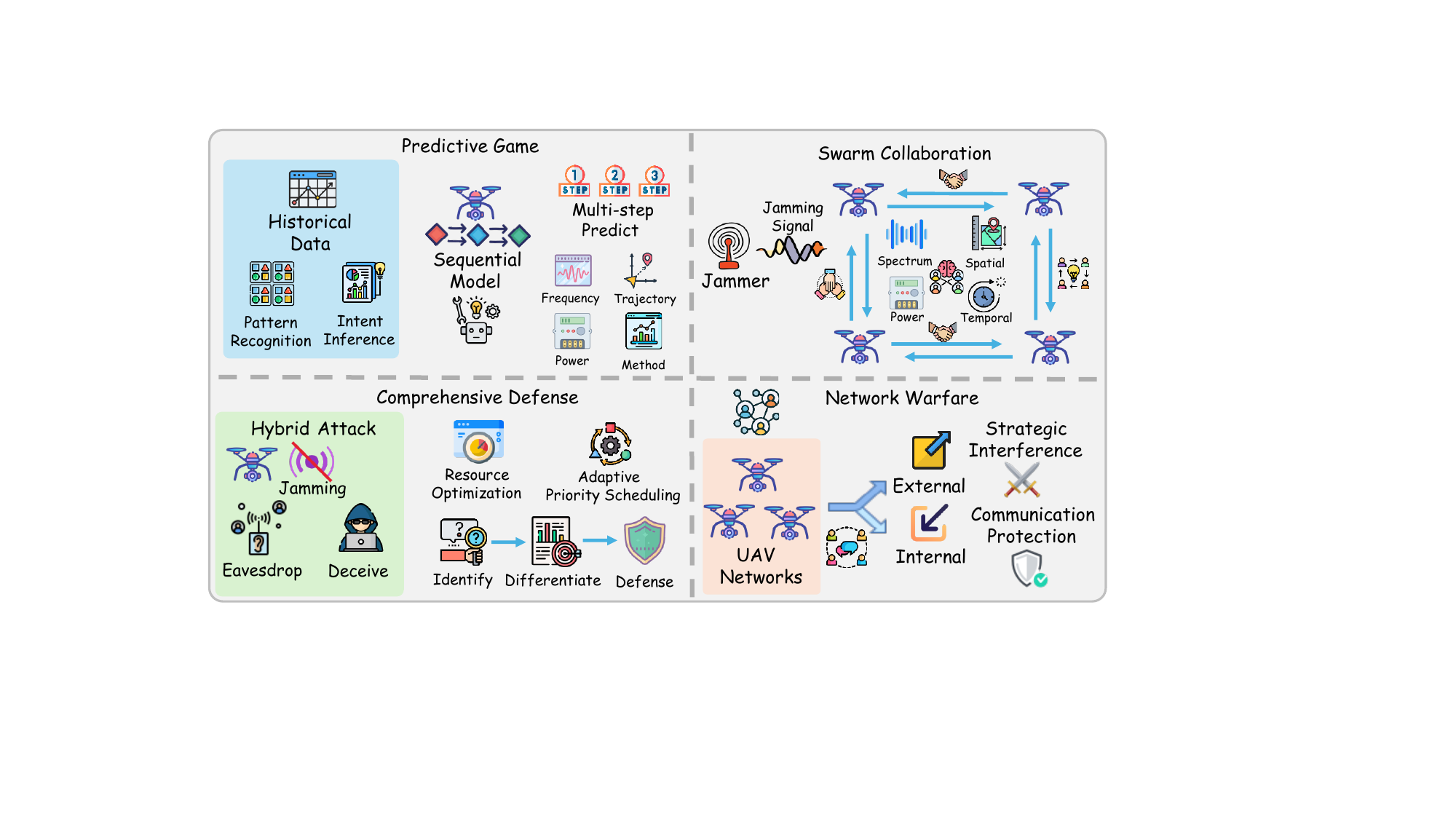}
    \caption{\textbf{The four future research directions for UAV Anti-Jamming technologies.}}
    \label{fig:four}
\end{figure*}

\section{Current Challenges and Future Research}
\subsection{Current Engineering Challenges}
Agent-based UAV anti-jamming communication faces several severe engineering challenges. A primary challenge is the highly dynamic and adversarial nature of the wireless environment, coupled with intelligent jamming, which places stringent demands on the real-time decision-making capabilities and adaptability of UAV anti-jamming agents. Furthermore, incomplete information and environmental uncertainty mean agents often must make decisions with partial or missing data, increasing learning difficulty and potentially degrading communication performance.

UAV systems are also hampered by the dual challenges of individual platform physical constraints and swarm coordination complexities. Individual UAVs have limited standalone anti-jamming capabilities due to factors like payload and computational power. While swarms can enhance resilience, they introduce intra-swarm interference, resource consumption, and challenges in distributed coordination within large-scale dynamic networks. Concurrently, the continuous evolution of intelligent jamming strategies challenges the defense system's ability to identify and adapt to novel attacks.

Additionally, deploying learning-based anti-jamming agents presents challenges with policy real-time applicability and convergence. Advanced reinforcement learning algorithms often require extensive training, conflicting with real-time adversarial needs. Enhancing learning efficiency, accelerating convergence, and ensuring real-time decision-making while maintaining policy effectiveness is a key engineering problem. The agent's robustness and generalization ability in diverse environments and against unknown jamming strategies are also critical for successful application.

\subsection{Future Research Directions in UAV Anti-Jamming}
Future UAV anti-jamming research must significantly enhance autonomous countermeasure capabilities by deepening intelligence in perception, decision-making, and execution. Key directions include: proactive threat mitigation using prediction and game-theoretic principles, collaborative anti-jamming and internal communication optimization for UAV swarms, comprehensive defense architectures resilient to hybrid attacks, and novel competitive strategies in network-centric adversarial environments(see Fig. \ref{fig:four}).

\textbf{Anti-jamming Techniques Integrating Prediction and Game Theory:}  This direction focuses on leveraging historical jamming data, employing sequential models for pattern recognition and intent inference, to develop efficient multi-step prediction models overcoming single-step limitations\cite{GCN}. The goal is to empower UAV agents to proactively execute evasive maneuvers such as frequency agility, power control, or trajectory adjustments based on predictions. Concurrently, strategies are needed to counter intelligent jammers, by enhancing the UAV's own perception, learning, and decision-making for rapid adaptation, and by exploiting potential limitations in the jammer's learning process.

\textbf{Collaborative Anti-Jamming and Internal Communication Optimization for UAV Swarms:}
The core research lies in developing collaborative anti-jamming strategies and internal communication optimization for swarms. This necessitates efficient distributed perception and information-sharing for unified situational awareness, and exploring multi-domain (power, spectrum, spatial, temporal) collaborative decision-making and resource optimization using methods like multi-agent reinforcement learning to counter external jamming. Concurrently, mechanisms such as distributed spectrum allocation, adaptive channel access, and power control are crucial for coordinating internal co-channel interference arising from swarm density and dynamics, alongside designing low-overhead, scalable interference management for reliable communication in large swarms.
 
\textbf{Comprehensive Defense Techniques Against Hybrid Attacks:} Addressing the increasing use of hybrid attacks integrating multiple techniques, for example, jamming, eavesdropping, and spoofing, future research must achieve breakthroughs in real-time fused perception and attribution of heterogeneous threats. This requires agents to accurately differentiate, identify, and correlate concurrent threats. Another key direction involves resource optimization and dynamic priority allocation for conflicting defense objectives, such as when facing both strong jamming and high-threat eavesdropping. This also includes exploring the dynamic generation and evaluation of multi-effect synergistic defense action sequences that achieve outcomes surpassing linear superposition.

\textbf{Competing Mobile Networks Anti-Jamming:} This approach centers on network-versus-network strategic confrontation, requiring UAV networks to integrate offense and defense by proactively executing strategic jamming against adversarial networks. Agents must model the enemy network, infer strategic intentions, and dynamically optimize their resources between internal communication protection and external strategic interference. The ultimate goal is to seize and maintain information superiority in multi-network adversarial environments by learning and executing complex competitive strategies.

\section{Conclusion}
This review provides a comprehensive survey of agent-based anti-jamming techniques for UAV communications. Against the backdrop of UAV communications facing severe challenges from heterogeneous interference sources—including malicious jammers, intra-swarm interactions, and natural environmental disturbances—this paper introduces the concept of an anti-jamming intelligent agent for UAV communications and systematically organizes the field around its core "Perception-Decision-Action" (P-D-A) closed-loop framework. Within this framework, the paper elaborates on key technical challenges and representative intelligent solutions across various stages, from environmental perception to intelligent decision-making and action execution. Special emphasis is placed on employing game theory to model the interactions between UAVs and jammers, as well as leveraging reinforcement learning and its variants to derive robust and adaptive anti-jamming strategies. Furthermore, the paper analyzes the practical advantages and potential limitations of current methodologies. Finally, it identifies engineering challenges in implementing anti-jamming intelligent agents and outlines promising future research directions. The continued advancement of these agent-based approaches is critical to ensuring reliable and resilient UAV communications in increasingly complex and adversarial electromagnetic environments.

\ifCLASSOPTIONcaptionsoff
  \newpage
\fi

\bibliographystyle{IEEEtran}
\bibliography{bibtex/IEEEexample.bib}


\end{document}